\documentstyle[bo99,epsfig]{article}

\title{Observing the effects of strong gravity with future X-ray missions}

\author{Christopher~S.~Reynolds\thanks{\ \ Hubble Fellow}}                                                       
\affil{JILA, University of Colorado, Campus Box 440, Boulder, CO~80309, USA}                                                
\begin{document}

\maketitle

\begin{abstract}
Spectroscopy of the broad iron iron with {\it ASCA} and {\it BeppoSAX}
has up opened the innermost regions of accreting black hole systems to
detailed study.  In this contribution, I discuss how observations with
future X-ray missions will extend these studies and all us to
observationally address issues which are currently only in the realm of
the theorists.  In particular, high-throughput spectroscopy with {\it
XMM} and, eventually, {\it Constellation-X} will allow the full
diagnostic power of iron line variability to be realized.  Instabilities
of the inner accretion flow, the geometry of the variable X-ray source,
and the black hole mass and spin will all be open to study.  Eventually,
X-ray interferometry will allow direct imaging of the black hole region
in nearby active galaxies, thereby providing the ultimate probe of black
hole astrophysics.
\end{abstract}

\section{Introduction}

As we have heard in this meeting, X-ray spectroscopy with {\it ASCA} and
{\it BeppoSAX} are providing probes of the region very close to the
supermassive black holes in active galactic nuclei (AGN).  In
particular, detailed observations and modeling of the broad $K\alpha$
fluorescence emission line of iron, which is thought to originate from
the surface layers of the inner accretion disk, allow us to probe the
inner disk structure and strong-field gravity in completely new ways.
The current observational status of this field has been summarized in
Dr.~Nandra's contribution in this volume.  In this paper, I will discuss
what there still is to learn, and how observations with future X-ray
missions will help us understand the environment near an accreting
supermassive black hole (SMBH).

As one might expect, the region close to an accreting SMBH is complex,
with many basic issues still unknown to us.  At a fundamental level, the
mass of most {\it active} SMBHs is very uncertain.  Furthermore, there
are essentially no robust indicators of black hole spin.  Many models
for the radio-loud/radio-quiet dichotomy of AGN postulate that the black
hole mass and, especially, the spin are the control parameters that
determine the radio-loudness of the object.  However, without
observational signatures of black hole masses and spins, it will be
difficult or impossible to test such models.  In additional, the physics
governing the interaction of the accreting matter with the SMBH is far
from clear.  Some of the outstanding questions are:
\begin{enumerate}
\item Does the inner accretion disk in some objects become hot
and geometrically-thick (see Dr. Sambruna's contribution in this volume
for a suggestion that this might be the case in broad line radio
galaxies)?
\item Is the violently variable X-ray emission due to magnetic flares on
the accretion disk surface, or changes within a central corona sitting
within the cold accretion disk?
\item What happens within the radius of marginal stability?   Does this
region have observational relevance?  For example, Krolik (1999)
recently suggested that the magnetic field becomes very strong in this
region, and as a result Alf\'en waves might plausibly transport
significant amounts of energy from this region into an inner corona or
the rest of the disk.
\item How are jets launched from the black hole region and collimated,
and what contribution do they make to the emissions observed from
non-blazar AGN.
\end{enumerate}
This article describes how future X-ray observations may attempt to
disentangle these phenomena.

\section{Current uncertainties and pure spectral studies}

The accretion disk model is highly successful at explaining the X-ray
reprocessing spectrum observed in many AGN.  A small number of AGN
(MCG--6-30-15, Tanaka et al. 1995; NGC~3516, Nandra et al. 1999;
NGC~4151, Wang et al. 1999) have been the subject of very long
integrations with {\it ASCA} yielding high quality iron line profiles
which match the predictions of the accretion disk model well (Fabian et
al. 1995).   However, there are still ambiguities present.

Firstly, a time-averaged iron line profile contains no information about
the mass of the central black hole.  All parameters relevant to
determining the line profile scale with the gravitational radius.
Secondly, and more interesting from an astrophysics point of view, the
line profile is sensitive to the X-ray source geometry, accretion disk
structure (including the region inside the innermost stable orbit), and
the spin of the SMBH.  Degeneracies exists in the sense that different
astrophysical assumptions and space-time geometries can produce very
similar iron line profiles.  The best studied example of this degeneracy
is the case of the very-broad state of the iron line in MCG--6-30-15
found by Iwasawa et al. (1996).  Making the standard assumptions that
the line emission is axisymmetric, and there is only emission from
outside of the radius of marginal stability, Iwasawa et al. (1996)
suggested that the SMBH in this object must be rapidly rotating to
produce a line as broad and redshifted as that seen.  Dabowski et
al. (1996) computed grids of iron line profiles for various values of
the SMBH spin with the same assumptions and set a formal limit of
$a>0.94$ on the spin of this SMBH.  However, Reynolds \& Begelman (1997)
showed that the same iron line profile can result from a non-rotating
SMBH if a high-latitude X-ray source illuminates disk material within
the radius of marginal stability.  This is an explicit demonstration of
how uncertainties in the assumed astrophysics (e.g. the X-ray source
geometry) leads to the degeneracy between models with very different
space-time geometries (i.e. Schwarzschild vs. extremal Kerr).  In a
rather different vain, Weaver \& Yaqoob (1998) showed that
non-axisymmetric obscuration of the line emitting region could also
reproduce these data.

\begin{figure}[t]
\centerline{\psfig{figure=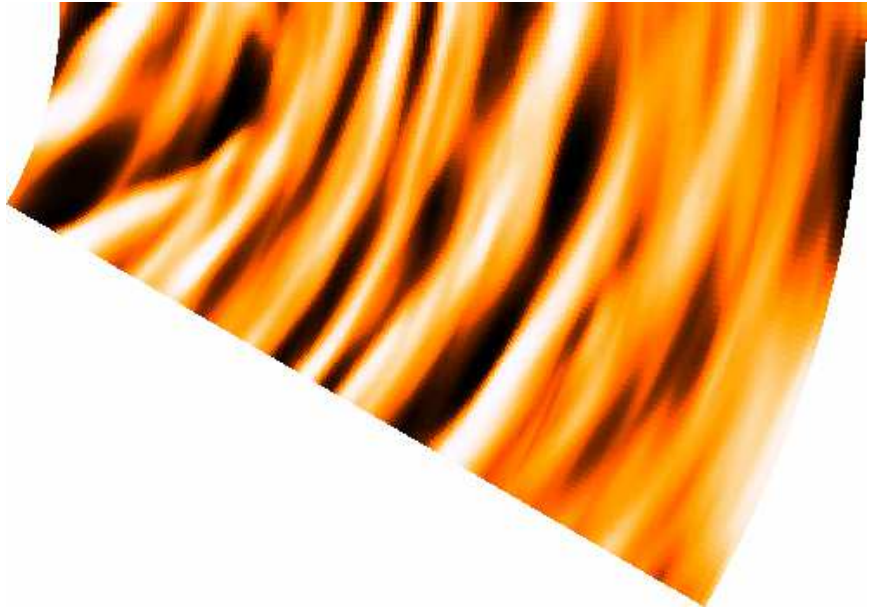,width=0.9\textwidth}}
\caption{Density structure in a slice through an MHD simulation of disk
accretion in a pseudo-Newtonian potential.  The inner edge of the
simulated wedge is at $r=4GM/c^2$ and the outer edge is at $r=12GM/c^2$.
Strong clumping can be seen at all radii, and especially within the
radius of marginal stability at $r=6CM/c^2$.  From Armitage \& Reynolds
(2000).}
\end{figure}

The first question to address is whether better spectroscopy with much
higher signal-to-noise and/or larger bandpass than {\it ASCA} and {\it
BeppoSAX} will remove these degeneracies.  Returning to the example of
MCG--6-30-15, Young, Fabian \& Ross (1998) showed that iron fluorescence
from material within the radius of marginal stability would be
accompanied by a large iron edge.  While it is questionable whether the
current {\it ASCA} data are of sufficient quality to rule out the
presence of such an edge, one might think that this would be a tell-tale
signature that could be used to distinguish the Schwarzschild and
extremal-Kerr models for this object.  However, it is important to
realize that such conclusions are at the mercy of extra epicycles of
astrophysical theory.  Both the Reynolds \& Begelman (1997) and Young et
al. (1998) models assume a smooth accretion flow within the radius of
marginal stability.  But strong magnetic fields in that region will
inevitably produce clumping of the material which will in turn lower the
ionization parameter of the material which produces the X-ray reflection
signatures (Armitage \& Reynolds 2000; also see Fig.~1).  This, in turn,
may diminish the depth of the iron edge that one would expect in the
spectrum.

\section{Iron line variability}

Spectral variability, and in particular variability of the broad iron
line, is a powerful probe of AGN central engines.  Many of the
degeneracies described above can be broken by considering line
variability.  In this section, I shall distinguish three types of line
variability and discuss how the study of each may help unravel the
complexities of these systems.

\subsection{Structural changes in the source}

\begin{figure}
\centerline{\psfig{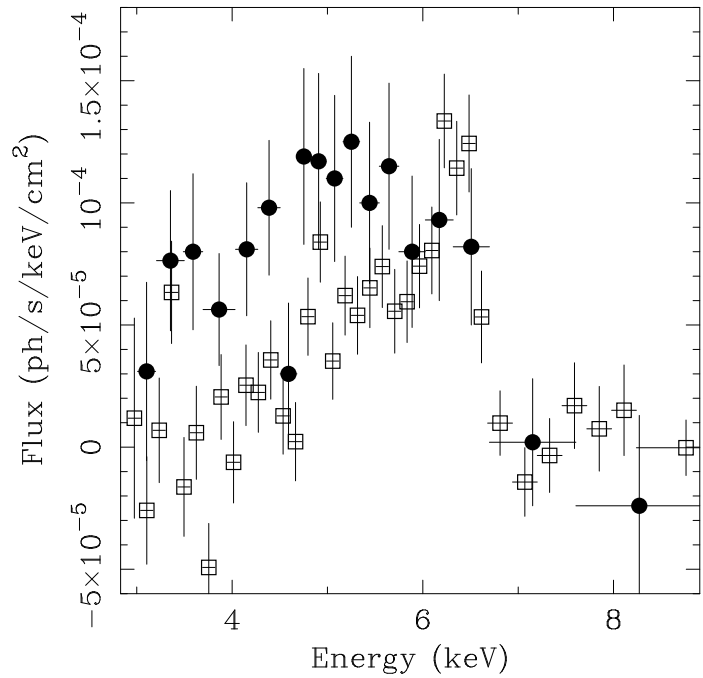}}
\caption{Iron line variability in MCG--6-30-15 detected by {\it ASCA} by
Iwasawa et al. (1996).  The open squares show the `normal' state of the
line whereas the filled circles show the `very-broad' state of the line,
during which time the continuum level was seen to drop dramatically.}
\end{figure}

As has already been mentioned above, {\it ASCA} has already seen broad
iron line variability in several objects, e.g. MCG--6-30-15 (Iwasawa et
al. 1996) and NGC~4051 (Wang et al. 1999).  Figure~2 shows the line
variability in MCG--6-30-15 in which the line changed from its `normal'
state (shown with open squares) to a very broad and strong state (shown
by filled circles).  This change in line profile accompanied a sharp
drop in the continuum flux level during an event that lasted at least
60\,ksec (which is greater than the dynamical timescale $t_{\rm dyn}$
for the inner accretion disk by a factor of $\sim 100$ or more for any
plausible SMBH mass; Reynolds 1999).  Unless the occultation scenario of
Weaver \& Yaqoob (1998) is correct, some dramatic change in the
structure of the accretion disk and/or the geometry of the illuminating
X-ray source is required to produce such dramatic and long-lived line
changes.  Changes in the thermal structure of the disk, which occur on a
timescale of $t_{\rm th}\sim t_{\rm dyn}/\alpha$ (where $\alpha\sim
0.1-0.01$ is the standard viscosity parameter), may produce this type of
variability.

Even given the long-lived nature of these events, {\it ASCA} cannot
produce high signal-to-noise line profiles in the different states.
This hampers our ability to probe details of the disk/corona variability
using these line changes.  {\it XMM} will completely change this
situation.  With an effective area at iron line energies more than a
factor of $10$ greater than {\it ASCA}, very high quality iron line
profiles will be obtained at different times as a source such as
MCG--6-30-15 undergoes one of these events.  While I dare not predict
what these observations will find, these studies will undoubtedly
revolutionize our understanding of the kind of instabilities suffered by
the inner accretion disk and X-ray emitting corona.

\subsection{Orbiting flares}

The X-ray emission from most AGN is observed to be highly variable on
timescales down to (our best estimate for) the dynamical timescale.
Whether the X-ray emission is due to magnetic flares exploding out of
the accretion disk or some other instability in a hot disk corona, the
instantaneous X-ray emission is likely to be non-axisymmetric.  If these
non-axisymmetric structures are long lived (i.e. survive at least a
couple of dynamical timescales), the iron line will be observed to
undergo distinct profile changes as the system orbits the central SMBH.

The computation of observables from an orbiting hot-spot on an accretion
disk around a black hole is a classical problem and has been worked on
my many authors (e.g. Ginzburg \& Ozernoi 1977, Bao et al. 1994, Bromley
et al. 1997).  Most recently, Ruszkowski (1999; also see contribution in
this volume) has computed the observed iron line variability when it is
powered by an X-ray flare that is co-rotating with the disk. {\it XMM}
should be able to track these profile changes and measure several key
parameters.  Firstly, the period and amplitude of energy variations in
the peak energy of the iron line are an easy and robust way of
determining the black hole mass.  Note that the inclination can be
measured from the time-averaged iron line profile and so is a
known quantity in this calculation.  Secondly, departures from
sinusoidal time-dependence of the iron line peak can be attributed to
relativistic effects and used to probe, for example, the spin parameter
of the black hole.  Such observations may yield signatures of a spinning
black hole: if iron line variations are found that imply a flare
orbiting on a circular orbit at a radius less the Schwarzschild radius
of marginal stability ($r=6GM/c^2$), a rapidly rotating black hole is
will be implied.

\subsection{Reverberation}

If some X-ray flares are very short lived, or activate rapidly (as
compared to the light-crossing time of the inner accretion disk), line
profile changes due to the finite speed of light will occur.  This then
raises the possibility of performing `reverberation mapping' of the
central regions of the SMBH accretion disk (Stella 1990; Reynolds et
al. 1999).    

In principal, reverberation provides powerful diagnostics of the
space-time geometry and the geometry of the X-ray source.  When
attempting to understand reverberation, the basic unit to consider is
the point-source transfer function, which gives the response of the
observed iron line to an X-ray flash at a given location.  As a starting
point, one could imagine studying the brightest flares in real AGN and
comparing the line variability to these point-source transfer functions
in an attempt to measure the SMBH mass, spin and the location of the
X-ray flare.  By studying such transfer functions, it is found that a
characteristic signature of rapidly rotating black holes is a `red-tail'
on the transfer function.  This feature corresponds to highly redshifted
and delayed line emission that originates from an inwardly moving ring
of illumination/fluorescence that asymptotically freezes at the horizon
(see Reynolds et al. 1999 for a discussion of this feature).

\begin{figure}
\hbox{
\psfig{figure=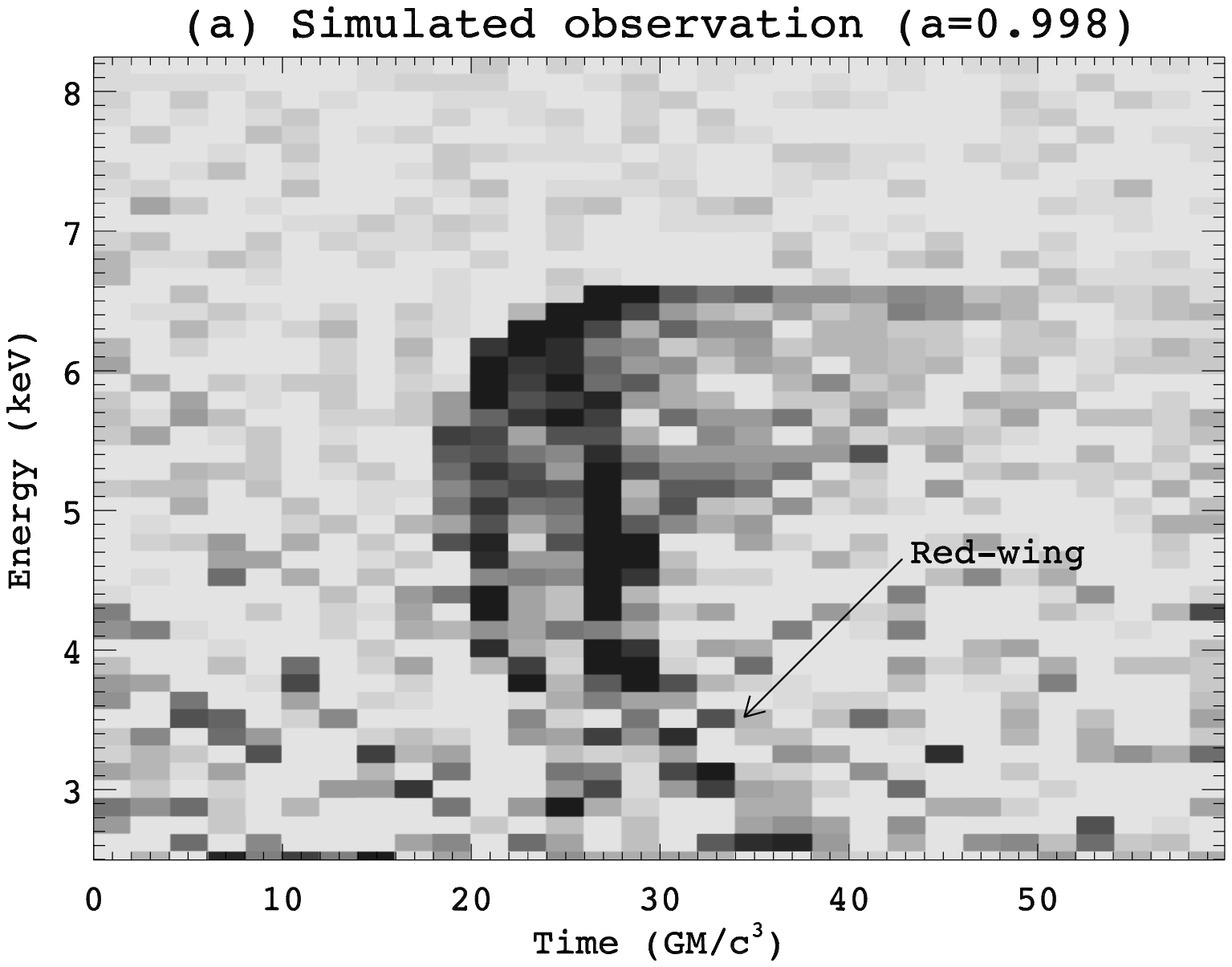,width=0.5\textwidth}
\psfig{figure=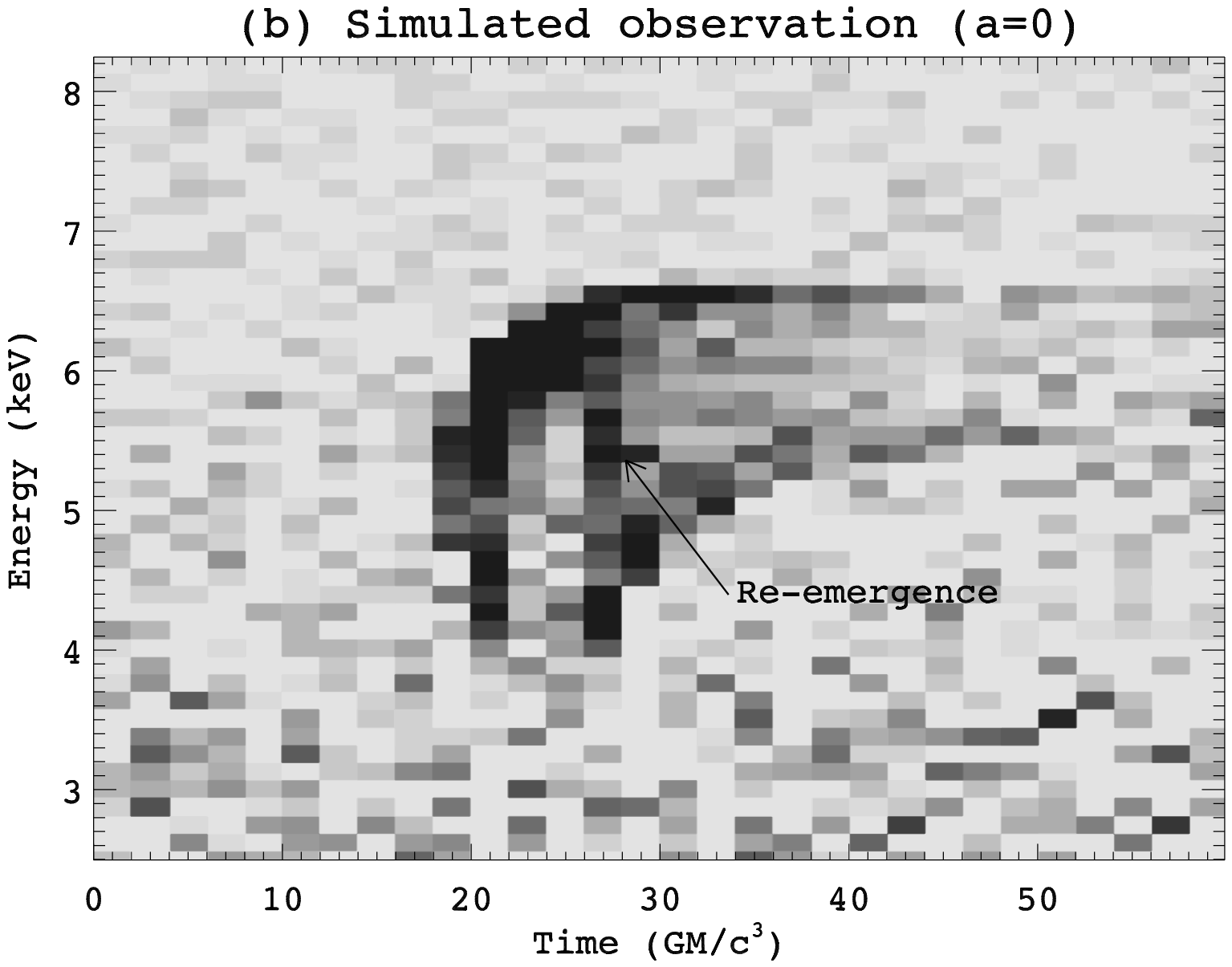,width=0.5\textwidth}
}
\caption{{\it Constellation-X} simulations of iron line reverberation.
Panel (a) shows the case of a rapidly rotating SMBH whereas panel (b)
shows a non-rotating SMBH.   In both cases, an X-ray flash on axis at a
height of $10GM/c^2$ has been assumed and the iron line response
calculated for an accretion disk inclination (away from normal) of
30$^\circ$.   Sequential 1000\,s {\it Constellation-X} observations of the
time varying iron line are then simulated, continuum subtracted, and
stacked in order to make an observed transfer function.  Figure from Young
\& Reynolds (1999).}
\end{figure}

The primary observational difficulty in characterizing iron line
reverberation will be obtaining the required signal-to-noise.  One must
be able to measure an iron line profile on a timescale of $t_{\rm
reverb}\sim GM/c^3\approx 500 M_8\,{\rm s}$, where we have normalized to
a mass of $10^8\,{\rm M}_\odot$.  This requires an instrument such as
{\it Constellation-X}.  Figure~3 shows that {\it Constellation-X} can
indeed detect reverberation from a bright AGN with a mass of $10^8\,{\rm
M}_\odot$.  Furthermore, the signatures of black hole spin may well be
within reach of {\it Constellation-X} (Young \& Reynolds 1999).
Although these simulations make the somewhat artificial assumption that
the X-ray flare is instantaneous and located on the axis of the system,
it provides encouragement that reverberation signatures may be
observable in the foreseeable future.

Of course, the occurrence of multiple, overlapping flares will also
hamper the interpretation of iron line reverberation.  The best way to
disentangle these flares is still the subject of current work.  However,
{\it Constellation-X} may have the required signal to noise to allow the
direct fitting of multiple transfer functions to real data (see Young \&
Reynolds 1999).

\section{Direct Imaging of black hole accretion disks}

\begin{figure}
\centerline{\psfig{figure=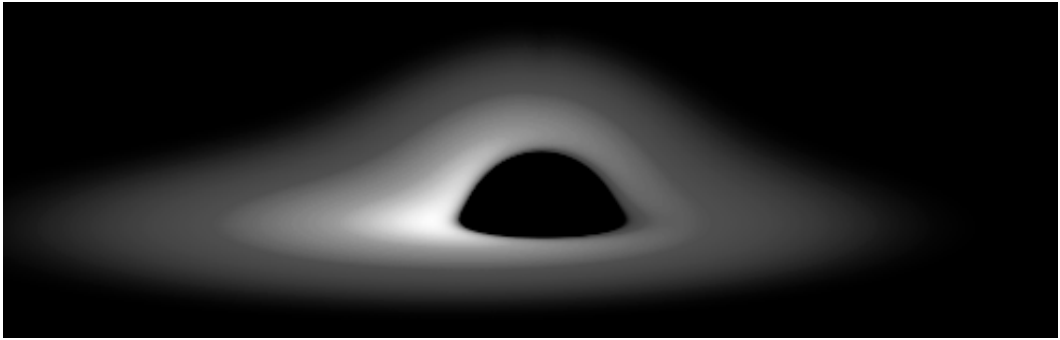,width=0.35\textwidth,angle=270}}
\caption{Theoretical image of a nearly edge-on accretion disk around a
Schwarzschild black hole.  The hole in the center of the image
corresponds to the radius of marginal stability at $r=6GM/c^2$.  The
distortions in the image of the far side of the accretion disk are due
to strong light bending effects.  In the future, X-ray Interferometry
will allow us to obtain such images for real systems.}
\end{figure}

I will end by briefly discussing an exciting idea which will allow us to
image the central regions of nearby AGN with sufficient angular
resolution to probe structure on scales smaller than the size of the
event horizon.  By combining diffraction limited X-ray optics with the
interferometric technologies that are currently being developed for the
Space Interferometer Mission ({\it SIM}), it is within our technological
reach to construct an X-ray interferometer capable of achieving
sub-microarcsecond resolution (this concept has become known as {\it
MAXIM}, the Micro-arcsec X-ray Interferometer Mission; see {\it
http://maxim.gsfc.nasa.gov/}).

As well as the obvious appeal of directly imaging an accreting black
hole, an observatory capable of achieving $0.1\mu\,{\rm arcsec}$
would yield major scientific return.  The geometry of the X-ray source
(and the spatial nature of the X-ray flares) would be open to direct
imaging studies.  X-ray activity or fluorescence from within the radius
of marginal stability could be easily seen (this region would be well
resolved).  We might also expect there to be X-ray emission from the
base of the jet in the region where the magnetic field couples to the
black hole spin via the Blandford-Znajek process.  Such emission could
be imaged, thereby providing the first look at these exotic physical
mechanisms at work.  If an interferometer can be constructed with
sufficient effective area, we will be able to use the fluorescent iron
line to make detailed velocity maps across the image.  These velocity
fields would provide direct constraints of the black hole mass and spin,
and implicitly provide a stringent test of strong field General
Relativity.

\section{Conclusions}

The immediate environment of an accreting supermassive black hole is
extremely exotic.  Broad iron lines provide us with the best tool to
date for studying these regions.  {\it ASCA} and {\it BeppoSAX}
observations have already shown us that the accretion disk in at least
some AGN extends very close to the black hole (and maybe so close as to
suggest that the black hole must be rotating).  Furthermore, the
detection of broad iron line variability by {\it ASCA} is most likely
tracking structural changes in the accretion disk and/or X-ray emitting
corona.  However, large effective area detectors are required to make
further progress.  {\it XMM} will allow these structural changes to be
characterized in detail, thereby probing the instabilities that affect
the inner accretion disk/corona.  Furthermore, {\it XMM} will allow us
to study iron line variability caused by the accretion disk rotation,
allowing us to measure the mass of the black hole and constrain the
location/lifetime of the X-ray flares.  Eventually, {\it
Constellation-X} will allow us to search for iron line reverberation.
The detection of reverberation will give robust signatures of black hole
spin and provide the tools to study the inner disk structure in
unprecedented detail.

Further in the future, direct imaging of the inner disk and black hole
region in nearby AGN will be possible using X-ray interferometry.  This
will provide the ultimate observational probe of black hole
astrophysics.

\begin{acknowledgements}
CSR appreciates support from Hubble Fellowship grant HF-01113.01-98A.
This grant was awarded by the Space Telescope Institute, which is
operated by the Association of Universities for Research in Astronomy,
Inc., for NASA under contract NAS 5-26555.
\end{acknowledgements}

\end{document}